\documentclass[%
 reprint,
superscriptaddress,
 amsmath,amssymb,
 aps,
prl,
]{revtex4-2}

\usepackage{graphicx}% Include figure files
\usepackage{dcolumn}% Align table columns on decimal point
\usepackage{bm}% bold math
\begin{document}

%\preprint{APS/123-QED}

\title{Antiferromagnetic behavior in self-bound one-dimensional composite bosons}

\author{M. C. Gordillo}
\email{cgorbar@upo.es}
\affiliation{Departamento de Sistemas F\'{\i}sicos, Qu\'{\i}micos y Naturales, Universidad Pablo de
Olavide, Carretera de Utrera km 1, E-41013 Sevilla, Spain}
\affiliation{Instituto Carlos I de F\'{\i}sica Te\' orica y Computacional, Universidad de Granada, E-18071 Granada, Spain.}

\date{\today}% It is always \today, today,

\begin{abstract}
The structure of self-bound one-dimensional droplets containing a mixture of Ytterbium fermionic isotopes ($^{173}$Yb, $^{171}$Yb) is calculated by means of a diffusion Monte Carlo technique.  
We considered only balanced setups in which all the atoms of one isotope are spin-polarized, while the atoms of the other can have up to three different spin values,  that difference being a necessary requirement to achieve stable systems. Our results indicate that these droplets consist of consecutive "molecules" made up of one $^{173}$Yb and one $^{171}$Yb atom.  In other words, we have up to three different kinds of composite bosons, corresponding to the number of spin components in the non-polarized isotope.  The fermionic nature of those Yb atoms makes pairs with identical spin composition avoid each other,  creating a Pauli-like-hole filed by another molecule in which at least one of the Yb atoms has a different spin from that of their closest neighbors.  This effective repulsion is akin to an antiferromagnetic short-range interaction between different kinds of composite bosons.      
\end{abstract}

\maketitle

%\section{Introduction}

The study of self-bound Bose-Bose droplets can be traced to a seminal paper by Petrov \cite{petrov2015} that suggested that binary bosonic arrangements with both attractive and repulsive interactions could avoid collapse due to purely quantum effects.  Petrov's conclusions gained substantial support through the experimental realization of these droplets \cite{cabrera,soliton,freedrop,dipolar1}, and originated an extensive body of theoretical work 
on such systems (see for instance Refs. \onlinecite{rep2020,frontiers2021} and those therein). 
An obvious follow-up to this line of research is the consideration of Bose-Fermi ensembles, already stable in the mean field picture \cite{sal1,njp2016,scipost2019,njp2019,peacock}. Recent experimental 
work attests to the very rich phase diagram of those Bose-Fermi composite systems \cite{nak40}.

The next logical step involves the exploration of Fermi droplets. Focusing specifically on cold atom systems with only  short-range attractive interactions in strictly one-dimensional (1D) environments, 
to have an stable droplet we would need atoms of at least two different spin types.  The problem is that  under such circumstances,  these atoms pair up to form bosonic molecules that do not interact with each other \cite{1dfermion,yoprr} due to the dual Pauli avoidance exhibited by fermions within a pair with respect to their counterparts with the same spin in other molecules.  So a set of balanced (with the same number) of spin-up and spin-down 1D fermions cannot be self-bound.  

In spite of that,  there is a way to produce 1D fermi self-bound drops. 
%In this work, as in previous literature, liquid means self-bound, i.e., with an energy smaller that that of the individual constituents. 
The recipe implies to have atoms with at least three different spin flavors, as
have been proved to work in small clusters with mixtures of fermionic Ytterbium ($^{173}$Yb and $^{171}$Yb) isotopes \cite{yoprr}.  
We will show that in those systems,  the attractive interactions between atoms of different isotopes produce as many kinds of bosonic molecules  as the different spin combinations  we can assemble to create them.  Thus,  the Pauli-like avoidance between identical composite bosons 
creates an effective antiferromagnetic interaction that depends exclusively on the nature of the units formed,  not being engineered via a external potential as in previously considered Bose systems \cite{antitheory,antiexp1,antiexp} by modifying their standard ferromagnetic coupling.   

The 1D clusters in this work will be described by the following Hamiltonian \cite{su6su2,su6su2ol,yoprr}: 
\begin{eqnarray}\label{hamiltonian}
 H=\sum_{i=1}^{N_p}\frac{-\hbar^2}{2m}\nabla_i^2 + g_{1D}^{173-171} \sum_{i=1}^{N_{173}} \sum_{j=1}^{N_{171}} \delta(x^{173}_i - x^{171}_j)
 \nonumber \\
+ g_{1D}^{173-173} \sum_{b>a} \sum_{i=1}^{n_{173,a}} \sum_{j=1}^{n_{173,b}} \delta(x^{173}_{a,i} - x^{173}_{b,j})
 \nonumber \\
+ g_{1D}^{171-171} \sum_{b>a} \sum_{i=1}^{n_{171,a}} \sum_{j=1}^{n_{171,b}} \delta(x^{171}_{a,i} - x^{171}_{b,j}),
\end{eqnarray}
where $N_p$ is the total number of fermions, while $N_{173}$ and $N_{171}$ are the total number of $^{173}$Yb and $^{171}$Yb atoms. 
In this work, $N_{173}$= $N_{171}$ = $N_p$/2.  $m$ is the mass of the atoms,  judged to be close enough in both isotopes to be described by a single parameter. 
No confining external potential in the $x$ direction was imposed. 
$n_{173,{ab}}$ and $n_{171,{ab}}$ are the number of atoms with spins $a$ and $b$. Identical fermions avoid each other by Pauli's exclusion principle. The $g_{1D}$ parameters depend on the 1D-scattering lengths, $a_{1D}$, via $g_{1D}^{\alpha,\beta} = -2 \hbar^2/m a_{1D}(\alpha,\beta)$,  with $a_{1D}$ defined by \cite{Olshanii}:
\begin{equation} \label{a1D}
a_{1D}(\alpha,\beta) = -\frac{\sigma_{\perp}^2}{a_{3D}(\alpha,\beta)} \left( 1 - A \frac{a_{3D}(\alpha,\beta)}{\sigma_{\perp}} \right),
\end{equation}
with A=1.0326 and $(\alpha,\beta)$ = $(173,171)$. $\sigma_{\perp} = \sqrt{\hbar/m \omega_{\perp}}$ is the  oscillator length  in the transversal direction, depending on the transversal confinement frequency, $\omega_{\perp}$, taken to be in the range 2$\pi\times$40-100 kHz, tight enough to produce a quasi-one dimensional system. $a_{3D}(\alpha,\beta)$ are the three-dimensional esperimental  scattering lengths between isotopes \cite{pra77}. The nature of the iteractions depends on the sign of those $a_{3D}$'s: atractive for the $^{173}$Yb-$^{171}$Yb and  $^{171}$Yb-$^{171}$Yb pairs and repulsive in the   $^{173}$Yb-$^{173}$Yb case.  %No confining-induced resonance is possible in the considered range of  parameters.  
                 
To solve the Schr\"odinger equation derived from the Hamiltonian in Eq. \ref{hamiltonian}, we used the  fixed-node diffusion Monte Carlo 
(FN-DMC) algorithm, that gives us the exact ground state of a 1D system of fermions \cite{hammond,ceperley} starting from an initial approximation to the exact wavefunction.  
Following Refs. \cite{su6su2,su6su2ol,yoprr},  we used:
\begin{eqnarray} \label{defa}
\Phi(x_1,\cdots,x_{N_p}) =
 \mathcal{A}[\phi(r_{11'}) \phi(r_{22'}) \cdots \phi(r_{N_{173},N_{171})}]
 \nonumber \\
\prod_{b>a}
\prod_{i=1}^{n_{173,a}} \prod_{j=1}^{n_{173,b}}  \frac{\psi (x_{a,i}^{173}-x_{b,j}^{173})}{(x_{a,i}^{173}-x_{b,j}^{173})} \nonumber \\
\prod_{b>a}
\prod_{i=1}^{n_{171,a}} \prod_{j=1}^{n_{171,b}}  \frac{\psi (x_{a,i}^{171}-x_{b,j}^{171})}{(x_{a,i}^{171}-x_{b,j}^{171})},
\end{eqnarray}
where $\mathcal{A}[\phi(r_{11'}) \phi(r_{22'}) \cdots \phi(r_{N_{173},N_{171})}] $ is the determinant of a square matrix
whose dimension is $N_{173} \times N_{171}$ \cite{pandaripande} and takes care of the interactions between pairs of particles of different isotopes 
separated a distance $r_{ij}$. The terms $(x_{a,i}^{\alpha}-x_{b,j}^{\alpha})$ in the denominator of Eq. \ref{defa} correct the spurious nodes 
between atoms of the same isotope with different spins (see Refs. \onlinecite{su6su2,yoprr} for further details). 
$\phi(r_{ij'})$'s are the solutions of the Sch\"odinger equation for a pair of 1D-particles
interacting via an attractive delta potential \cite{griffiths}: 
\begin{equation} \label{solution1D}
\phi(|x_i^{173}-x_{j'}^{171}|) = \exp \left[-\frac{|g_{1D}^{173,171}|}{2} |x_i^{173}-x_{j'}^{171}| \right].
\end{equation}
On the other hand, 
$\psi (x_{a,i}^{\alpha}-x_{b,j}^{\alpha})$'s are Jastrow functions that introduce the correlations
between pairs of particles of the same isotope belonging to different spin species $a,b$.
For the repulsively interacting $^{173}$Yb-$^{173}$Yb pair,  we have \cite{gregoritesis}:
%\begin{widetext}
\begin{equation}\label{delta}
\psi (x_{a,i}^{173}-x_{b,j}^{173}) = \cos(k[|x_{a,i}^{173}-x_{b,j}^{173}|-R_m])
\end{equation}
%\end{widetext}
when the distance between atoms, $|x_{a,i}^{173}-x_{b,j}^{173}|$, was smaller than a variationally obtained parameter, $R_m$, and 1 otherwise, $k$ being the solution of the transcendental equation $k a_{1D}(173,173)\tan(k R_m)=1$. When the pair of particles of the same isotope attract each other, as in the $^{171}$Yb-$^{171}$Yb case, the Jastrow has
the form of Eq.  \ref{solution1D} \cite{gregoritesis,su6su2},
but with a different value of the defining constant, $g_{1D}^{171,171}$. 

\begin{figure}
\begin{center}
\includegraphics[width=0.8\linewidth]{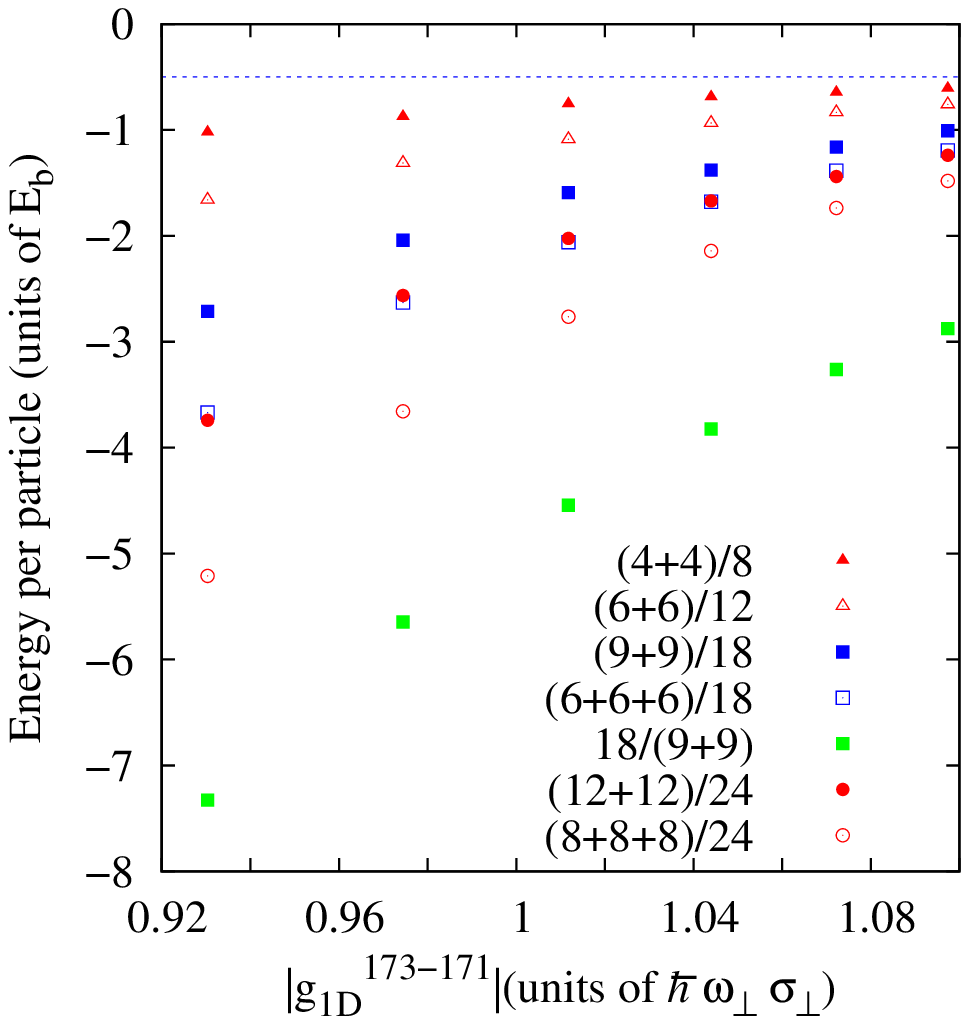}
\caption{Energy per Yb atom in units of $E_b$ for balanced clusters with different compositions. The clusters are named as
$(\sum_{a=1}^{s_{173}} n_{173,a}) / (\sum_{b=1}^{s_{171}} n_{171,b})$i, with $s$ the number of spin types. The error bars are of the size of the symbols and not shown for simplicity. The dashed 
line correspond to a set of non-interacting $^{173}$Yb-$^{171}$Yb molecules. 
}
\label{fig1}
\end{center}
\end{figure}

In this work, we considered 1D arrangements with $N_p$ in the range 16-48. This means $N_{171}$ = $N_{173}$ going from
8 to 24.  We have chosen clusters that contains up to three different spin types for $^{173}$Yb (out of 6 available), and up to two for $^{171}$Yb (all the possible types for this fermionic isotope).  In all cases, one of the Yb isotopes, the majority component, was spin-polarized. That those clusters are stable irrespectively of their spin composition can be shown in Fig. \ref{fig1}: the energy per atom is displayed as a function of
$|g_{1D}^{173-171}|$, that depends on the transverse confinement via Eq. \ref{a1D}.  We can see that the energy per particle
is always lower (more negative) than the corresponding to a set of not interacting bosonic pairs,  $E_b$/2 =  $(g_{1D}^{173,171})^2/(8 \hbar \omega_{\perp} \sigma_{\perp})$ \cite{griffiths}.  This is due to the relaxation in the Pauli's restrictions between atoms of the same isotope and different spins belonging to different Ytterbium molecules.  We can see also that when the spin-polarized atom is $^{173}$Yb, the clusters are more stable,  due to the residual attraction between the spin-up and spin-down atoms of the $^{171}$Yb.  

\begin{figure}
\begin{center}
\includegraphics[width=0.8\linewidth]{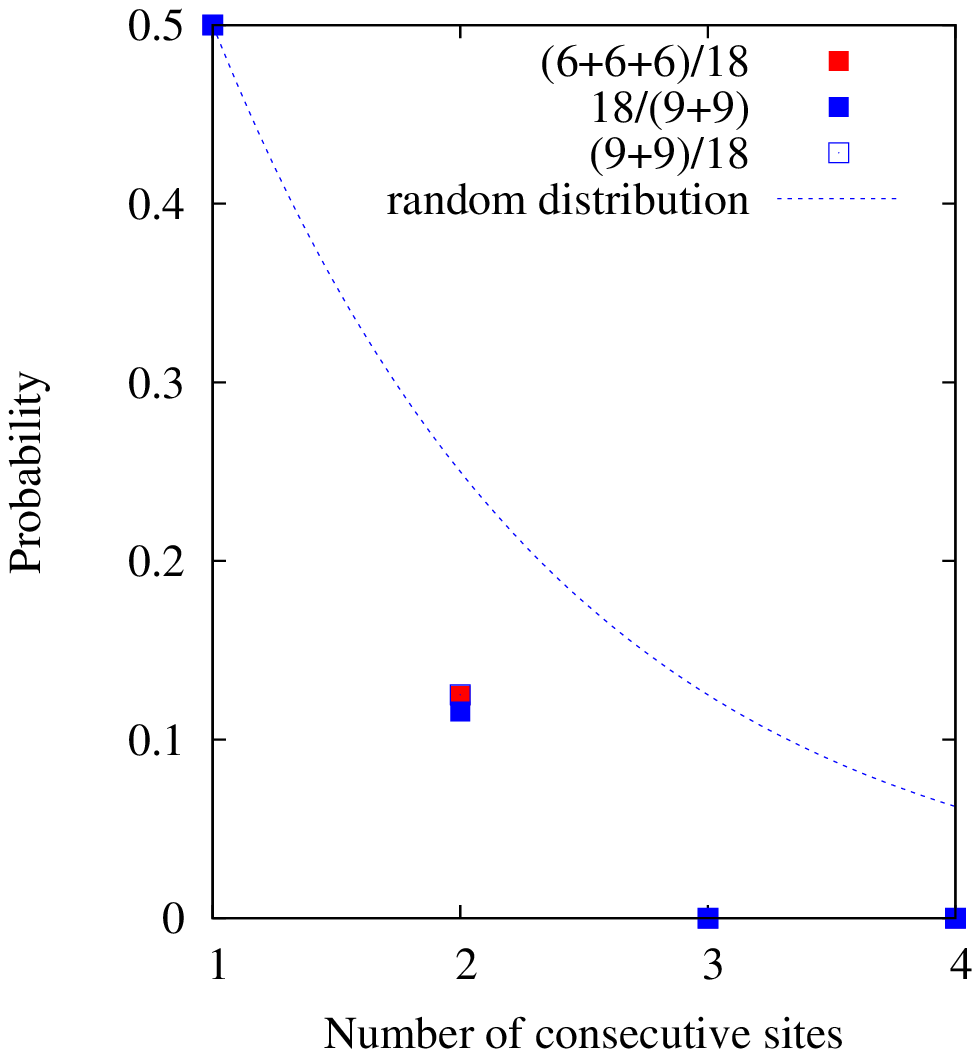}
\caption{Probability of having $j$ consecutive sites occupied by atoms of the majority component for clusters of different compositions. Symbols,  simulation results;  dotted line, 
values for those probabilities in a random distribution. Error bars are of the size of the symbols and not shown for simplicity.
}
\label{fig2}
\end{center}
\end{figure}

The next step will be to study the actual  distribution of the Yb atoms in those 1D ensembles.  To do so,  we started by assigning each atom an index, $i$,  that corresponds to its relative position in the collection of atoms that form the cluster., i.e.,  for the atom with the lowest $x$ coordinate we have $i$=1,  the one with the second-to-lowest $x$ gets 
$i$=2 and $i$= $N_p$ for the Yb atom at the end of the 1D row.   After that, we introduce the variable $n_i$,  that takes the value 1 if the atom with index $i$ belongs to the majority  (spin polarized) component and 0 otherwise.  With that parameter, we can build the correlators:
%\begin{equation}
%c_1= \sum_{i=1}^{N_p} n_i/N_p,  \nonumber
%\end{equation}
%\begin{equation}
%c_2 = \sum_{i=1}^{N_p-1} n_i n_{i+1}/(N_p-1)  \nonumber 
%\end{equation}
%\begin{equation}
%c_3 = \sum_{i=1}^{N_p-2} n_i n_{i+1} n_{i+2}/(N_p-2)  \nonumber
%\end{equation}
%\begin{equation}
%c_4 = \sum_{i=1}^{N_p-3} n_i n_{i+1} n_{i+2} n_{i+3}/(N_p-3) 
%\end{equation}
\begin{equation}
c_j = \sum_{i=1}^{N_p-j+1} \prod_{k=0}^{j-1} \frac{n_{i+k}}{N_p-j+1} 
\end{equation}
to characterize the atom ordering inside the cluster.  $c_j$ can be defined as the probability of having a set of $j$ consecutive majority component atoms.  For instance,  $c_4$ is the average of the product $n_i n_{i+1} n_{i+2} n_{i+3}$ for all the possible values of $i$, divided by the total number of sets of four consecutive positions,  $N_p-3$.  
The results are given in Fig.  \ref{fig2} for $c_1$ to $c_4$,  $c_1$ being simply the ratio of the majority to the total number of atoms,  i.e., 0.5 for all the arrangements considered in this work.  We represent there different examples of clusters with $N_p$ = 36,  but the results for other arrangements are virtually identical to those shown. 

From Fig. \ref{fig2} we can see that the distribution of atoms is far from random and independent of the cluster composition.  In particular, it is impossible to have more than two $^{171}$Yb (in the first example) close together. 
In addition,  the fact that $c_2 \ne$ 0 also implies that we do not have a perfect ordering in which each $^{171}$Yb is necessarily followed by a $^{173}$Yb atom and vice versa.   On the other hand,
if we calculate the same corrrelator,  but only for $i$=1 (not considering the entire sum,  only the first term),  we find $c'_2$ = 0 and the same happens when $i$=$N_p-1$.  Moreover,  if we calculate $c'_2$=0 for atoms other than in the majority component,  we get also $c'_2$ for $i$ =1,$N_p-1$, i.e., the first two (and the last two) atoms belong to different isotopes.  All of this necesarily implies that the atoms distribute themselves in consecutive $^{173}$Yb-$^{171}$Yb units with different orientations.  
\begin{figure}
\begin{center}
\includegraphics[width=0.8\linewidth]{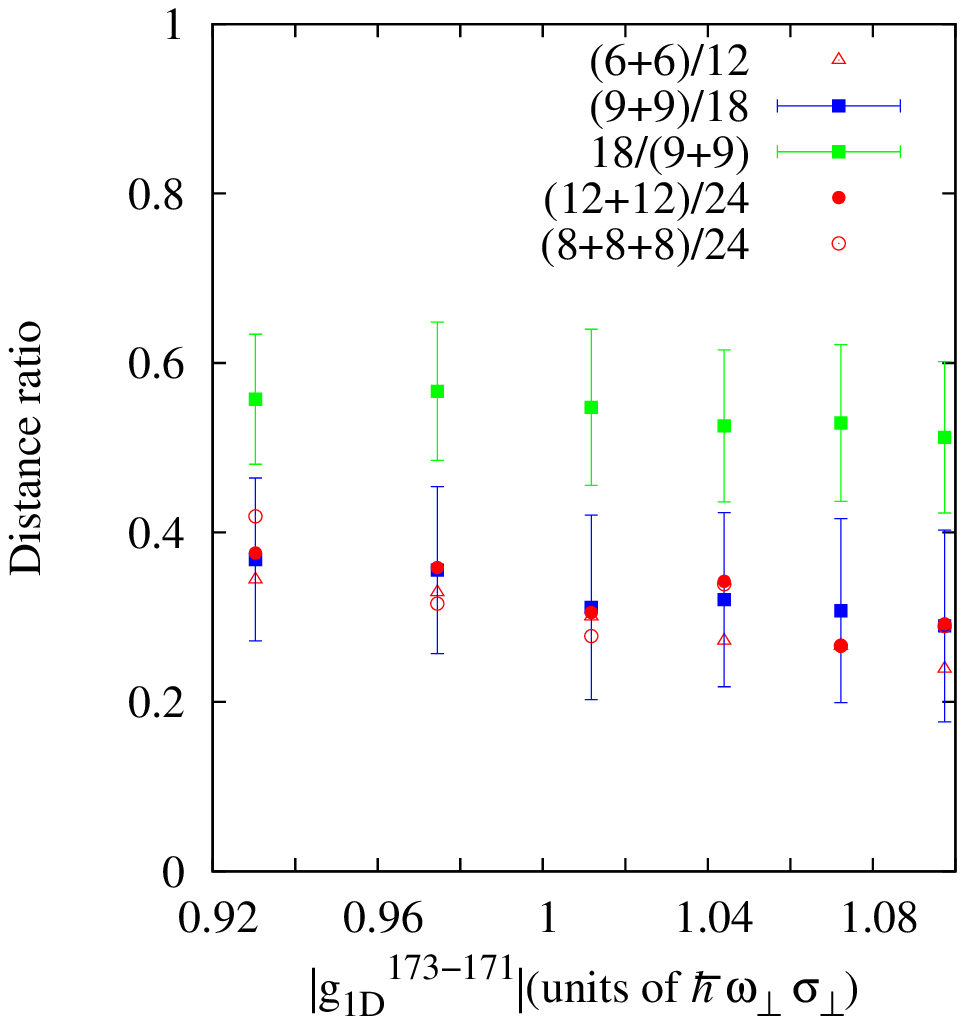}
\caption{Ratio between the distances between consecutive ($i$, $i+1$) and ($i+1$,$i+2$) pairs with $i$ odd for clusters of different compositions as a function of the $|g_{1D}^{173-171}|$ parameter.  See further explanation in the text.
}
\label{fig3}
\end{center}
\end{figure}

Unfortunately, this does not necessarily mean that the atoms will form $^{173}$Yb-$^{171}$Yb (in whatever order) molecules in which the atoms in the pair are closer together than a couple of atoms belonging to different units. 
%Since the $c'_2$ correlator implies that the first pair of atoms in any cluster should belong to different isotopes,  
To check whether we really have those couples,   we have calculated the ratio of the distances for all the consecutive ($i$,$i+1$) and ($i+1$,$i+2$) pairs with $i$ odd.  This means that we divided the distance between the first and second atom in a cluster by the distance between the second and third
%,  the distance between the third and forth by the one between the forth and fifth and 
and so on.  The average of those ratios should be smaller than one if ($i$,$i+1$) pairs with $i$ odd are formed.  This is exactly what we see in Fig. \ref{fig3} for clusters
of different sizes and compositions.  Thus, it is safe to say that atoms of different isotopes always pair together to form composite bosons well separated from their  adjacent couplets.   We can see also that the relative distance between separate molecules depends on the nature of the spin-polarized isotope and not on the size of the cluster nor on the number of spin types.  Reasonably enough, we also observe that when the interactions between non-polarized components are attractive, the distance between molecules decreases and  the ratio displayed in Fig. \ref{fig3} increases, as in the 18/(9+9) cluster.

\begin{figure}
\begin{center}
\includegraphics[width=0.8\linewidth]{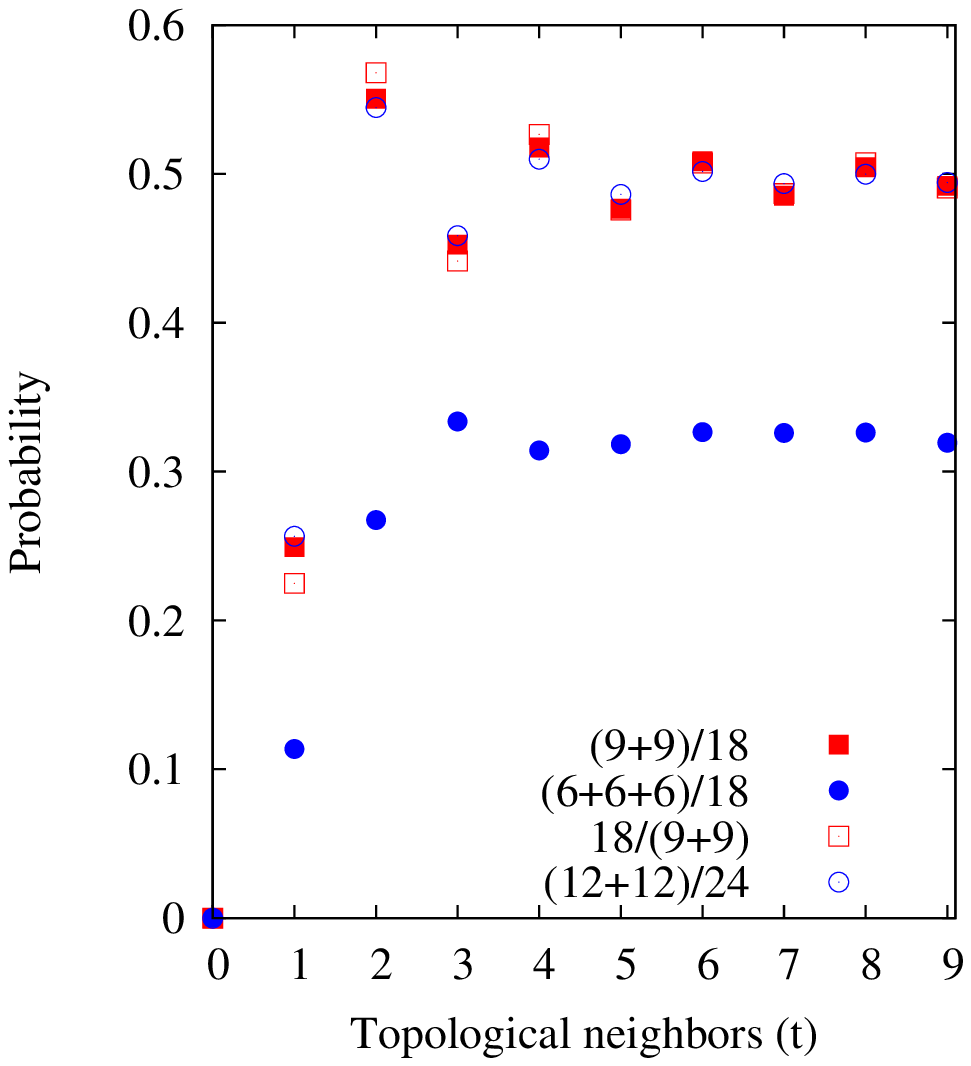}
\caption{Probability of having an identical molecule as a $t$-th topological neighbor for clusters of different sizes and compositions.   Error bars are of the size of the symbols and not displayed by clarity.
}
\label{fig4}
\end{center}
\end{figure}

So far, we have stablished that all the clusters considered here are made up of molecules consisting of atoms of different isotopes.  However,  we do not know how those molecules distribute themselves inside the clusters.  To address that question, we calculated the probability of having two identical molecules at different topological distances.  So,  in  Fig.  \ref{fig4}, nearest neighbors correspond to an abscissa equal to one ($t$=1),  next-to nearest neighbors to two ($t$=2), and so on.  In a random distribution, that probability is the ratio between the number of pairs of a particular kind to the total number of pairs in the cluster.  This means 1/2 or 1/3 depending on the number of pair types.  In Fig. \ref{fig4}, we can see that the simulation results for several clusters are very close to those values for $t\geq$4.  However,  the deviations from the random limits are evident for closer neighbors.  In particular, there is a "hole" in the probability of having two identical nearest neighbors.  This decrease extends to $t$=2 for clusters with three kinds of molecules.  On the other hand,  it is more probable to have a next-to-nearest neighbor of the same type than what it would correspond to a random pair ordering.  The same can be said of the third neighbor for the (6+6+6)/18 case.   This is exactly the same  to what happens in all-$^{173}$Yb fermionic clusters \cite{su6jordi} and has its origin in the antiferrromagnetic interactions 
characteristic of systems of fermions.   We have then a set of composite bosons with short-range ($t\leq$4) effective antiferromagnetic interactions.  

Summarizing,  we have characterized the behavior of strictly 1D clusters of fermionic mixtures of Yb atoms, mixtures that have already been produced experimentally \cite{taie,pra103}. SInce the pairing of two fermions produces a boson,  the 1D arrangements considered in this work are composite bosons, but with properties derived from their fermionic constituents.  In particular,  they exhibit  a short-range avoidance of identical bosonic molecules. 
This is completely at odds with the standard ferromagnetic behavior one would expect from a set of regular bosons \cite{antitheory}.  In addition,  this is an intrinsic behavior that does not have to be engineered via an external optical potential as in Refs. \onlinecite{antiexp1,antiexp} and it should be,  therefore,  stable for long periods of time.

\begin{acknowledgments}

We acknowledge financial support from Ministerio de
Ciencia e Innovación MCIN/AEI/10.13039/501100011033 
(Spain) under Grant No. PID2020-113565GB-C22
and from Junta de Andaluc\'{\i}a group PAIDI-205.  
We also acknowledge the use of the C3UPO computer facilities at the Universidad
Pablo de Olavide.
\end{acknowledgments}

\bibliography{antiferro}% Produces the bibliography via BibTeX.

\end{document}